# A New Method Towards Speech Files Local Features Investigation

Rustam Latypov
Department of System Analysis and Information Technology
Kazan Federal University
Kazan, Russia
Roustam.Latypov@kpfu.ru

Evgeni Stolov
Department of System Analysis and Information Technology
Kazan Federal University
Kazan, Russia
ystolov@kpfu.ru.

**ABSTRACT**

There are a few reasons for the recent increased interest in the study of local features of speech files. It is stated that many essential features of the speaker language used can appear in the form of the speech signal. The traditional instruments — short Fourier transform, wavelet transform, Hadamard transforms, autocorrelation, and the like can detect not all particular properties of the language. In this paper, we suggest a new approach to the exploration of such properties. The source signal is approximated by a new one that has its values taken from a finite set. Then we construct a new sequence of vectors of a fixed size on the base of those approximations. Examination of the distribution of the produced vectors provides a new method for a description of speech files local characteristics. Finally, the developed technique is applied to the problem of the automatic distinguishing of two known languages used in speech files. For this purpose, a simple neural net is consumed.

**KEYWORDS**

Speech files, Local properties, Language distinguishing, Neural net

## 1 Introduction

The information technologies give a possibility meaningfully cutusage keyboard for making requests by changing the keyboard with a microphone. If a query is addressed to a person, there is no problem in recognition of the used request language. Now, consider a case there are two or more official languages in the country. A citizen forms a speech request, utilizing the preferred language, to a state body. As a rule, a request cannot be answered immediately, and the authority wants a printed version of the request. Various services can solve this problem, but before converting the query to a print version, the system must determine one of the possible languages used in the request [1].

There are many publications dedicated to the mentioned challenge. In [2], a review of the earlier implemented methods is presented. A set of phonemes corresponding to each of the considered languages is taken. The information system compares the current phoneme from the speech file with the ones installed in the set. The system determines the language in the input file, using the standard likelihood procedure. There are visible drawbacks to this approach. The templates of the chosen phonemes depend on the speakers. Thus, many templates from different speakers should be presented in the system. An option of the described procedure is comparing signals spectra in place of comparing the signals in the time domain [3]. More recently, neural networks based language identification systems (LID) approaches are becoming increasingly popular and reported to offer superior performance compared to traditional LID techniques [4]. Lately, end-to-end multilingual systems based on the sequence-to-sequence architecture are proposed [5]. A part of language parameters can be used for a unified architecture with a shared vocabulary among multiple languages. Recently, compelling systems developed which can automatically recognize speech narrated in one of the known languages [6], [7]. Those systems require significant resources, and only a restricted number of public authorities can afford such a system.

If we deal with two languages that have different origins, there is a chance to use some unique features in files that afford to determine used language without vast calculation. That is precisely the case we investigate in this paper: distinguish Tatar and Russian speech files.

It is known that languages differ in voice onset time (VOT) [8], [9]. The works, exploiting this peculiarity in files, suggest various methods for localization of consonants in speech. And then to measure the distance from the end of the consonant to the beginning of vocalization [10]. The principal obstacle is automatically obtaining the moment where vowel begins. It can be performed via analysis of the spectrum of the signal inside of the window, moving along file [11]. Another approach to the problem is based on the supposition that some sounds produced by native speakers have some peculiarities. Revealing such peculiarities, one can distinguish the languages. It was shown in [12] that the waveform corresponding to the consonants could be used as a label for solving the problem. A parametric curve approximates the wave, and the parameters providing the best approximation determine the language with high reliability. The form of the wave also defines some other features of the speech. In [13], it was shown that signal samples in the region of extreme value determine the instantaneous signal frequency. Hence these set some unique features of the speaker.

In this paper, we continue the investigation of the local properties of the signal. A kind of correlation must be presented among neighboring samples in speech. The standard procedures can not reveal that correlation since the signal referred to the part of the file corresponding to the consonant can not be viewed as a stationary signal. To overcome that obstacle, we use the following procedure. We move a window along with the signal. Then we approximate the signal inside the window with the values from a particular set $S$ and change the signal inside the window with a vector with items in $S$. After that, we train a neural net to restore the original signal via the vector. This net is used as a filter for investigation of the local properties of the file. That determines the language used by the speaker.

## 2   Approximation of signal

Let a discrete source signal be a sequence $Sign = <a_0, a_1, \ldots, a_{N-1}>$. We are going to replace the original signal with a stepwise sequence $Appr = <b_0, b_1, \ldots, b_{N-1}>$. In this article, we consider the case when a stepwise sequence has 5 levels. One can also consider the procedure for approximating the original sequence by a stepwise sequence with an arbitrary odd number of levels. First, we define four thresholds: $Th_0 < Th_1 < Th_2 < Th_3$. The rule for matching the original signal $a$ into level $b$ is defined by (1):

$$b = \begin{cases} 2 & if\ a \geq Th\_3 \\ 1 & if\ a \geq Th_2\ \&\ a < Th_3 \\ 0 & if\ a \geq Th_1\ \&\ a < Th_2 \\ -1 & if\ a \geq Th_0\ \&\ a < Th_1 \\ -2 & otherwise \end{cases} \quad (1)$$

Since we need a kind of optimal approximation, we have to choose the metric to compare approximation $Appr$ with the original signal $Sign$. As usual, we use the standard SNR to this end (2):

$$SNR(Signal, Appr) = 10 \cdot log_{10}\left(\frac{\sigma^2(Sign)}{\sigma^2(Sign-Appr)}\right) \quad (2)$$

where both the array are supposed to have equal standard deviations. If we go the standard way, then to get the optimal values for the thresholds in metrics (2), it is necessary to implement the optimization procedure with four variables. Instead, for optimization, we leverage suboptimal procedures that lead to acceptable results. We implement two modifications of the suboptimal algorithm. The source signal $Sign = Sign_+ + Sign_-$ where the first term contains non-negative and the second term – non-positive items. Each of the terms is approximated separately. We set a relation between the thresholds, that reduce the optimization procedure to the tabulation of a single parameter with a step. We suppose that $Th_2 > 0, Th_1 < 0$ and set $Th_3 = 2 \cdot Th_2$ and $Th_0 = 2 \cdot Th_1$. The suboptimal value for $Th_2$ is calculated by the procedure $getOptThresh$ where $In = Sign_+$. The value of the $Th_1$ is obtained by means of the same procedure where $In = -Sign_-$ and $Th_1$ equals the inverse value returned by the procedure. Let

$$Dset = \{-2, -1, 0, 1, 2\}. \quad (3)$$

The final approximation $Appr$ with items in $Dset$ of the source $Sign$ is calculated by the formula

$$Appr = getAppr(Sign_+, Thr_2) - getAppr(-Sign_-, -Thr_1).$$

An example of algorithmic approximation to a source signal is shown in Fig. 1.

---
**Procedure** : getOptThresh(In,StartThresh,Step)
1: $Snr, BestThresh \leftarrow -100, 0$ {Save best value of SNR and Threshold}
2: $\bar{In} \leftarrow In$ {Normalize by standard deviation}
3: $Thresh \leftarrow StartThresh$
4: **while** $Thresh < Max(\bar{In})$ **do**
5: $\quad Fl \leftarrow getAppr(\bar{In}, Tresh)$
6: $\quad \bar{Fl} \leftarrow Fl$ {Normalize by standard deviation}
7: $\quad Val \leftarrow SNR(\bar{In}, \bar{Fl})$ { Implement (2)}
8: $\quad$ **if** $Val > Snr$ **then**
9: $\quad\quad Snr, BestThresh \leftarrow Val, Thresh$
10: $\quad$ **end if**
11: $\quad Thresh \leftarrow Thresh + Step$
12: **end while**
13: **return** $BestThrash$

---

## 3   Vector presentation of stepwise signal

Our goal is to study the dependency of neighbor samples in speech files. Since supposition about stationary of the signal is not fulfilled, the standard correlation does not suit to this end. Our approach to the problem is as follows. We approximate the source signal using the described method and convert it into a sequence $Dseq$ with elements from the set $Dset$ (3). Then we choose a window of length $LW$, which slides along $Dseq$. Any position of the window gives a vector of length $LW$ containing items inside of the window. Note that any element of the resulting vectors is in $Dset$. We represent the vector

$$Vec = c_0, c_1, \ldots, c_{LW-1}$$

by an integer

$$I_{Vec} = \sum_{k=0}^{LW-1} c_k \cdot 5^k \quad (4)$$

Thus, we established a one-to-one correspondence between the constructed vectors and the set of integers in the interval

$$\left[-\frac{5^{LW}-1}{2}, \frac{5^{LW}-1}{2}\right].$$

```
Procedure : getAppr(In, Thresh)
1: Ln ← In {Length of signal}
2: Out ← Ln {Create zero array of length Ln}
3: Out ← 1 {Fill in the array with ones}
4: Indx ← where (In ≥ 2 · Thresh)
5: Out[Indx] ← 2
6: Indx ← where (In < Thresh)
7: Out[Indx] ← 0
8: return Out
```

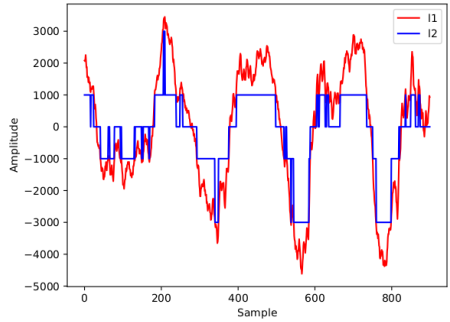

**Fig. 1. Source signal and its approximation with a five lelvel step-function. i1 -- source signal, i2 --- approximation**

Now, we can implement standard tools for the study of digital sequences. Let us create a histogram of all digits corresponding to the vectors of length *LW* (see *Algorithm* 1). The source file contains speech recorded at a frequency of 44100Hz. We display this simple algorithm consciously since the results of its work look unexpectedly (see Figs. 2, 3, 4). One can see that the histograms related to speech signals differ very slightly, whereas the interval of the values (5), used by the creation of the histograms, depends upon the length *LW* of the window.

```
Algorithm 1 Histogram of vector approximation of signal
Input:  Appr, LW
Output: BinValues
1: AllVal {All digital forms of vectors are collected here}
2: for I = 0 to length(Appr) − LW do
3:    Vec ← Appr[I : I + LW]{Interval of file}
4:    Ivec ← Vec {Implement (4)}
5:    AllVal[I]leftarrowIvec
6: end for
7: Bins ← histogram(AllVal, 30) {30 bins}
8: BinValues ← Bins/sum(Bins)
9: return BinValues
```

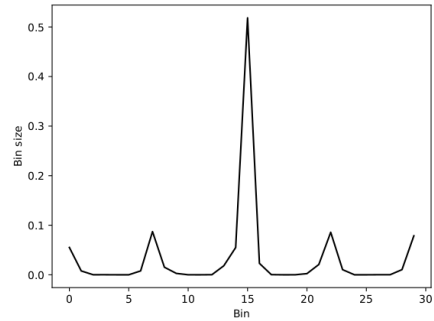

**Fig 2. Distribution of values. Russian female speaker, length of window equals 7**

On the other hand, the histogram in Fig. 5, corresponding to white noise, is distinguished from the previous pictures. The similarity of the histograms describing the distribution of values in speech files mirrors the existence of correlation among neighbor samples in such records, whereas that correlation is missing in white noise. Now, we refine the distribution of single vectors since the histogram describes **the integral situation. We c**ount how many times each vector, converted to digital form according to (4), is presented in the records utilized above. A result of the counting is shown in Fig. 6. Here, instead of the number of appearances of a vector in the record, the relative frequency is plotted on the Y-axis, so that we could compare data extracted from records of different durations. The results of analogous counting for white noise are depicted in Fig. 7. One can distinguish these pictures visually. The vectors having maximal frequencies in Fig. 6 are vectors with constant items related to the digits: $0 \to <0,0,0,0,0,0,0>$ , $39062 \to <2,2,2,2,2,2,2>$ , $1931 \to <1,1,1,1,1,1,1>$ and also the vectors with inverse components. We can hypothesize why structures of the histograms and the distributions of singe vectors with maximal frequency are independent of the length *LW* of the window used for the creation of the vectors. As before, let *Appr* be a file produced by the approximation procedure. Suppose that $Appr$ contains a series of constant signals $a, a, a, ..., a$ length $M$. We see that the vectors with invariable components have the maximal frequencies of appearance. While the window of the length $LW$ slides along $Appr$, a sequence of $M - LW + 1$ vectors with constant components arises. If we increase the length of the window by $K$, then the number of the created vectors decreases by $K$ elements. If the series is long enough, then the increase of the length of the window does not change the structure of the histogram. Fig. 8 supports this hypothesis. We collected all intervals of the kind $< 1,1, ...,1 >$ and sorted them using the length of the interval as a key in decreasing order (the first interval has the maximal length). We see that intervals of significant size are presented in the collection. This procedure was performed with two vectors with records of Russian and Tatar speech.

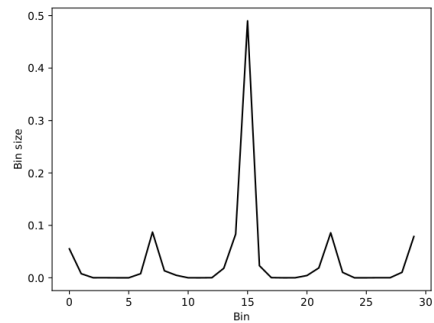

**Fig 3. Distribution of values. Russian female speaker. Length of window equals 11**

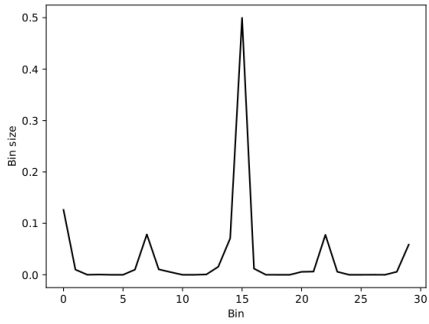

**Fig. 4. Distribution of values. Russian female speaker. Length of window equals 11**

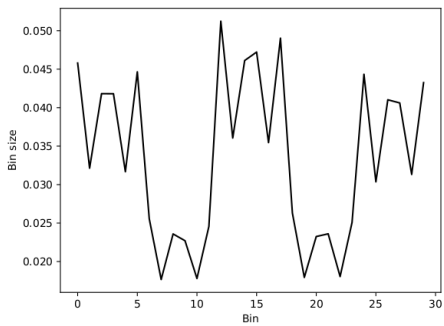

**Fig.5. Distribution of values. White noise. Length of window equals 7**

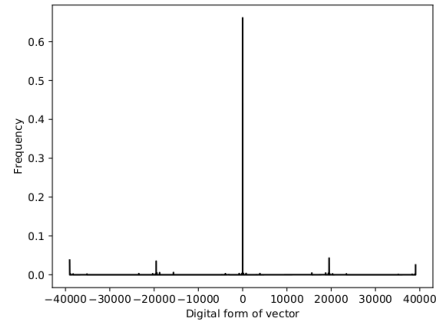

Fig.6. Relative frequences of various vectors in record. Russian male voice. Length of window equals 7

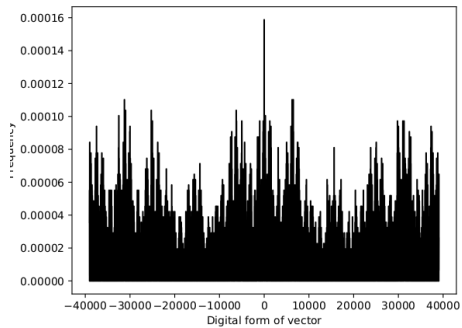

**Fig. 7. Distribution of vectors in windows of length 7 sliding along white noise lile**

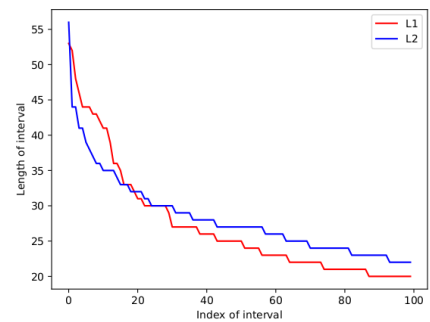

**Fig. 8. Vectors consisting of ones are sorted by their length. L1 -- Russian speaker, L2 -- Tatar speaker**

## 4   Distribution of vectors and recognition of the language used in records

The main goal of the research in this section is a test of the following hypothesis. In essence, the language of a record can be recognized on the base of the distribution of vectors if it is known in advance that only one of two languages is used. Our basic tool is the distribution of the vectors among various frequencies, which is shown in Fig. 6. In our case, the database has a limited size, since it is created by direct recording during television news. It contains 31 files with a Russian male voice, 36 files with a Tatar male voice, 26 files with a Russian female voice, and 24 files with a Tatar female voice. All files are recorded at a frequency of 44100Hz and have a duration of 6 to 12 sec.

## 4.1 The average vector

The simplest feature of the arbitrary distribution is the average vector. It is calculated according

$$Average = \sum_{Vec, I_{vec}>0} fr(Vec) \cdot Vec. \quad (6)$$

Here the summation is carried over all vectors having positive $I_{Vec}$ calculated via (4), $and\ fr(Vec)$ is the relative frequency of vector appearance. While summation, we limited ourselves with items having positive digital form because the graph in Fig. 6 is close to symmetric one. For any vector $Vec$, there exists inverse to its vector $-Vec$, and $fr(Vec) \approx fr(-Vec)$. It means that Since the size of our database is very small, we implement bootstrapping while calculation [14]. Algorithm 2 shows the main steps. For any file in the database, the corresponding average vector is obtained, and all vectors related to the same object (Russian male voice and the like) are collected in the matrix. The function $shuffle$ from the package $Numpy$ [15] permutes rows at any step. Half of the rows in any matrix are used for training. The rest of the rows are utilized for testing recognition of the language via the average vector.

**Algorithm 2** Bootstrap and neural net for language recognition

**Input:** $Matr1, Matr2, NumbOfTests$ {Matrices containing average vectors for compare and the number of bootstrap steps}
**Output:** $Numb1, Numb2$ {Numbers of correct recognized languages}
1: $Row1, Row2 \leftarrow Matr1, Matr2$ {Numbers of rows in matrices}
2: **for** $I = 1$ to $NumbOfTests$ **do**
3:    $Matr1Sh \leftarrow Matr1[shuffle]$ {Permutation of rows of $Matr1$ by shuffle function}
4:    $Matr2Sh \leftarrow Matr2[shuffle]$
5:    $Matr1ShL \leftarrow Matr1Sh[: Row1/2]$ {Half of the matrix used in learning procedure}
6:    $Matr2ShL \leftarrow Matr2Sh[: Row2/2]$
7:    $Matr \leftarrow Matr1Sh, Matr2Sh$ {Concatenation of two matrices}
8:    $TrainData \leftarrow Row1/2, Row2/2$
9:    $NeurNet \leftarrow Matr, TrainData$ {Training neural net}
10:   $Numb1 \leftarrow NeurNet, Matr1Sh[Row1/2 :]$ {Rest of the matrix used for recogniton}
11:   $Numb2 \leftarrow NeurNet, Matr2Sh[Row2/2 :]$
12:   **print** $Numb1, Numb2$
13: **end for**

## 4.1 The average vector

We used a simple neural net for clustering the average vectors depending on the used language. The net is created by means of the *Tensorflow-Keras* package [16], [17]. The net has two hidden layers with 5 neurons and $relu$ activation functions each, output layer having single neuron with $sigmoid$ activation function. The gained results are placed in Tables 1 and 2.

**Table 1. Results of language recognition.** $LW = 11$

| Correct recognized Russian and Tatar male voices ||
|---|---|
| Russian, total 16 | Tatar, total 18 |
| 16 | 12 |
| 11 | 12 |
| 8 | 11 |
| 16 | 10 |
| 16 | 12 |

**Table 2. Results of language recognition**

| Correct recognized Russian and Tatar female voices ||
|---|---|
| Russian, total 13 | Tatar, total 12 |
| 13 | 9 |
| 13 | 12 |
| 11 | 12 |
| 12 | 9 |
| 12 | 10 |

The obtained results can be viewed as encouraging, taking into account the minimal resources used by calculations.

## ACKNOWLEDGMENTS

This work was supported in part by the Russian Government Program of Competitive Growth of Kazan Federal University.